\def\({\left(}
\def\[{\left[}
\def\l\{{\left\{}
\def\){\right)}
\def\]{\right]}
\def\r\}{\right\}}
\def\what{\widehat}
\def\raw{\rightarrow}
\def\bA{\bar A}
\def\am{\bar A_{\rm max}}
\def\cf{{\cal F}}
\newcommand{\pd}[2]{\frac{\partial #1}{\partial #2}}
\newcommand{\A}[1]{\bmath{#1}}
\newcommand{\HALF}{\frac{1}{2}}

\def\etal{{\sl et al.} }
\def\ratio#1#2{{{#1}\over{#2}}}
\def\KH{Kelvin-Helmholtz }
\def\3d{three-dimensional }
\def\2d{two-dimensional }
\def\be{\begin{equation}}
\def\ee{\end{equation}}
\def\bef{\begin{figure}}
\def\ef{\end{figure}}

\def\DXDYCZ#1#2#3{\left({\partial#1\over\partial#2}\right)_{#3}}
\def\lapp{\mathbin{\raise2pt \hbox{$<$} \hskip-9pt \lower4pt
\hbox{$\sim$}}}
\def\gapp{\mathbin{\raise2pt \hbox{$>$} \hskip-9pt \lower4pt
\hbox{$\sim$}}}

\documentclass{aa}

\usepackage{txfonts}
\usepackage{graphicx}

\begin{document}

   \title{Time-dependent MHD shocks and line intensity ratios in the HH 30 jet:
A focus on cooling function and numerical resolution}

  \author{
  O. Te\c{s}ileanu\inst{1,2} 
\and S. Massaglia\inst{2} \and A. Mignone\inst{2} \and G. Bodo\inst{3} \and F. Bacciotti\inst{4}
}

\authorrunning{Te\c{s}ileanu et al.}

\titlerunning{Time-dependent MHD shocks and line ratios in the HH 30 jet}

\institute{
RCAPA - Department of Physics, University of Bucharest,
CP MG-11, RO-077125, Bucure\c{s}ti-M\~{a}gurele, Romania \and
Dipartimento di Fisica Generale dell'Universit\`a,
Via Pietro Giuria 1, I-10125 Torino, Italy\\
email: tesilean@ph.unito.it, massaglia@ph.unito.it, mignone@ph.unito.it, \and
INAF - Osservatorio Astronomico di Torino, Viale Osservatorio 20, I-10025
Pino Torinese, Italy\\
email: bodo@to.astro.it \and
INAF - Osservatorio Astrofisico di Arcetri, Largo E. Fermi 5, I-50125
Firenze, Italy\\
email: fran@arcetri.astro.it 
 }

   \date{Received; accepted }

\abstract{
The coupling between time-dependent, multidimensional MHD numerical 
codes and radiative line emission is of  utmost importance in the
studies of the interplay between dynamical and radiative processes in many 
astrophysical environments, with particular interest for problems involving radiative
shocks. 
There is a widespread consensus that line emitting knots observed in Herbig-Haro jets
can be interpreted as radiative shocks. Velocity perturbations at the jet base
steepen into shocks to emit the observed spectra. 
To derive the observable characteristics of the emitted spectra, such as line intensity ratios,
one has to study physical processes that involve the solution of the MHD equations coupled 
with radiative cooling in non-equilibrium conditions. 
}{
In this paper we address two different aspects relevant to
the time-dependent calculations of the line intensity ratios of forbidden transitions, resulting from the 
excitation by planar, time-dependent radiative shocks traveling in a stratified medium. 
The first one concerns the impact of the radiation and ionization processes included in the cooling model,
and the second one the effects of the numerical grid resolution.
}{
Dealing with both dynamical and radiative processes in the same numerical
scheme means to treat phenomena characterized by different time and length
scales. This may be especially arduous and computationally
expensive when discontinuities are involved, such as in the case of shocks.
Adaptive Mesh Refinement (AMR) methods have been introduced in order to alleviate 
these difficulties. In this paper we apply the AMR methodology to the treatment of radiating shocks 
and show how this method is able to vastly reduce the integration time.
}{
The technique is applied to the knots of the HH 30 jet to obtain the observed line intensity ratios and
derive the physical parameters, such as density, temperature and ionization fraction. We consider 
the impact of two different cooling functions and different grid resolutions on the results.
}{
We conclude that the use of different cooling routines has effects on
results whose weight depends upon the line ratio considered. Moreover, we 
find the minimum numerical resolution of the simulation grid behind the shock to achieve convergence
in the results. This is crucial for the forthcoming 2D calculations of radiative shocks.
}


\maketitle

\section{Introduction}
Supersonic flows are ubiquitous in the Universe: expanding supernova remnants, stellar
winds, AGN and Herbig-Haro jets are some examples, and shocks are abundant 
and prominent in these flows. Shocks located in extragalactic environments, such as AGN jets,
can be considered adiabatic, since the cooling time for thermal emission typically exceeds  by far
the source lifetime. On the other hand, shocks that form in galactic objects such 
as supernova remnants and Herbig-Haro jets must be treated including radiative effects.

Radiative shocks have been studied in steady-state conditions by several authors (e.g. Cox \&
Raymond \cite{CR85}, Hartigan et al. \cite{HA94}), who derived
the one-dimensional post-shock behavior of various physical parameters (temperature, 
ionization fraction, electron density, etc.), as 
functions of the distance from the shock front. From these
studies it turns out that the physical parameters vary behind the shock
front on scale lengths that differ by orders of magnitude, 
and this becomes a major problem when tackling the 
time-dependent evolution of such radiative shocks.
For example, in Herbig-Haro jets, one needs to treat in the same scheme
spatial scales well below $\sim 10^{13}$ cm, 
to reproduce the time-dependent post-shock temperature 
variations correctly, and scales $\lapp \ 10^{15}$ cm to study the behavior of the 
electron density (Massaglia et al. \cite{MA05a}, Paper I).
Under these conditions and with these requirements, employing an Adaptive Mesh Refinement (AMR)
technique becomes almost mandatory. This technique can
provide adequate spatial and temporal resolution by dynamically
adapting the grid to the moving regions of interest, giving
a tremendous saving in computational time 
and memory overhead with respect to the more traditional uniform grid approach.  

The important issue of numerical spatial resolution in 
time-dependent simulations of radiative shocks in 2D has been considered by \cite{Raga07}
employing an AMR technique as well. They discussed the dependence of the morphological structure of the
perturbations depending on the grid resolution. Their maximum grid resolution was with a cell
size corresponding to $1.5 \times 10^{12}$ cm. They found that while the detailed structure of the
shocks depends on the resolution, the emission line luminosities, integrated over the volume, are less dependent
on cell size. 

A second crucial question is: does the dependence on the cooling function details have qualitative
or quantitative effects on the calculated distributions of the physical parameters? Te\c{s}ileanu et al. (\cite{TA08})
have developed a detailed treatment including an ionization network of 29 species and compared,
at constant numerical resolution,
the resulting distribution of physical parameters (such as temperature, density, etc.) with the one obtained employing the
simplified cooling of Paper I, for a 2D cylindrical jet affected by perturbations that evolved in internal working surfaces.
They found quantitative differences but qualitatively the outcomes were very similar.

A more challenging problem than the determination of the integrated line
emissions and distributions of physical parameters is the calculation of line intensity ratios between
different species. In this case, the fractional abundances of the various species vary
in very different ways with space behind the shocks (see Paper I). This variation is mainly governed
by the temperature profile, which is typically a very steep function of space. Therefore we will examine
the effects of both the details of the cooling function and of the grid resolution adopted.
 
As discussed in Paper I and Massaglia et al. (\cite{MA05b}), observations of some HH jets 
at distances of a few arcseconds from the source typically show that the behavior of temperature, 
ionization and density along the jet is incompatible with a freely cooling jet.
It is now generally accepted that the line emission in HH jets is a result of the observation of several 
post-shock regions of high excitation with a filling factor of about 1\%. 
In our calculations, we will refer, in particular, to the observational results 
described in Bacciotti et al. (\cite{BE99}), where  
a specialized diagnostic technique has been applied to 
HST data of the HH 30 jet. 
Hartigan \& Morse (\cite{HM07}) also re-examined the 
HH 30 case using slitless spectroscopy performed with the Hubble Space Telescope
Imaging Spectrograph. Their results are fully 
consistent with the findings in Bacciotti et al. (\cite{BE99}).

In this paper, we will follow the dynamical evolution of an initial perturbation as it steepens into a (radiative) 
shock traveling along the jet, and derive the post-shock physical parameters consistently. 
From these parameters, we construct the synthetic theoretical emission line ratios to 
be compared with observations.
Using this setup, we implicitly consider that the medium in the real jet returns close 
to the steady jet conditions between the propagating shocks. This was verified in parallel
2D simulations (Te\c{s}ileanu et al. \cite{TA09}), as the 1D simulation, being done 
in the reference frame of the mean flow, lacks the steady flow of the jet. 
After the shock and the high-density post-shock zone pass, there follows an underdense zone, 
and then the medium returns to the stationary conditions of the constant flow. The distance 
after which this happens increases during the evolution of the shock from $0.5\times 10^{14}$cm 
close to the jet base to about $10^{15}$cm at 500AU from the base. This increasing distance 
becomes resolved observationally at about 200AU from the base of the jet, and also at such 
distances interactions between different shockwaves become likely, these being possible 
explanations for the oscillations visible in the observational data after 200AU. 
The emission at each point of the jet is 
computed by simulating the propagation of a shock produced at the base of the jet up to
that point.
As discussed before, we will adopt two different treatments for the radiative losses, namely:
i) the simplified cooling employed in Paper I and ii) the new cooling treatment described 
in Te\c{s}ileanu et al. (\cite{TA08}).

The plan of the paper is the following: In Section 2 we present the computation setup and the 
adopted techniques to model the problem; in Section 3 we apply the model to the case of HH 30 and
discuss the results obtained with the two cooling functions and the effects 
of the grid resolution; the conclusions are drawn in Section 4.

\section{The model}
%
%
%
%
%
In this Section
we give  a general outline of the model and the form of the initial perturbation.
More details on the setup can be found in Paper I, where we analyzed the case of 
the DGTau jet.

\begin{figure}
\resizebox{\hsize}{!}{\includegraphics{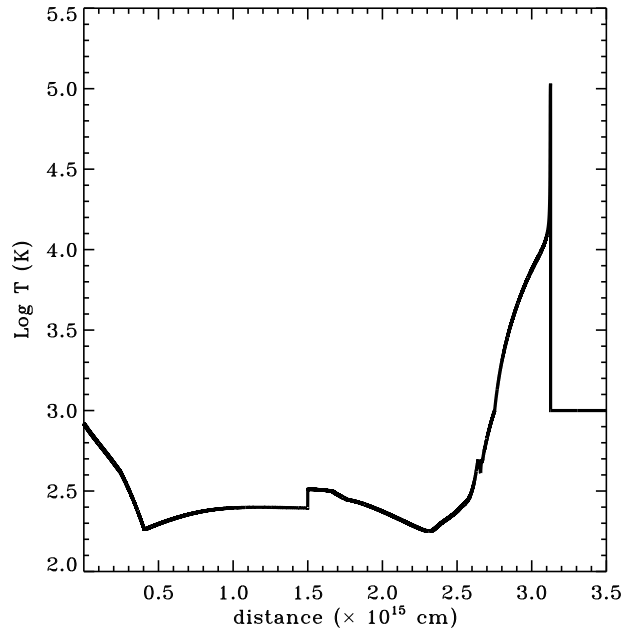}}
\caption{Snapshot of the temperature behavior vs distance for the 
         uniform grid, for a propagating shock in a stratified medium.}
\label{fig:unif}
\end{figure}

\begin{figure}
\resizebox{\hsize}{!}{\includegraphics{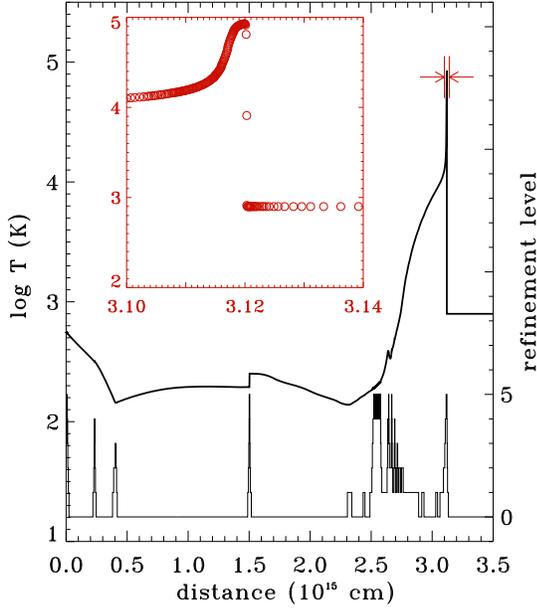}}
\caption{The same as in Fig. \ref{fig:unif} but for the AMR 
         integration and five levels of refinement, for an 
         equivalent resolution of 49152 cells, with a 
         zoom on the shock region. The histogram in the lower
         part of the plot (with the right vertical scale) 
         represents the refinement level employed in each cell
         of the simulation.}
\label{fig:amr}
\end{figure}

As mentioned in the Introduction, in the preshock conditions
assumed here (decreasing density departing from a few 10$^4$cm$^{-3}$, 
uniform temperature of 2\,000K) 
the post-shock temperature drops by about one order of magnitude
over a scale length lower than $\sim 10^{13}$cm i.e. a much smaller 
distance than the post-shock evolution that goes on over 
distances of a few times $10^{15}$ cm. This has been a 
compelling reason for developing and employing Adaptive Mesh Refinement, 
as in Paper I, described in detail in a companion paper (Mignone et al. \cite{MA09}).
We have verified the validity of this approach by computing 
several of the models presented in the next section on uniform 
and adaptive meshes and found a tremendous gain in efficiency. 
Figs. (\ref{fig:unif}) and (\ref{fig:amr}) show the results of 
one particular model computed on a uniform grid employing 
$49152$ zones and on an equivalent adaptive grid with $5$ levels 
of refinement (base level resolution is $1536$ zones).
On a Pentium IV processor with a 1.7GHz clock and 1 GB of RAM, 
the uniform grid approach took about 
$6.04 \times 10^4$ sec, while the AMR computation required only $238$ sec, up to
the same final integration time $t = 5\times 10^3$. 


\subsection{Model equations and initial perturbation}

We restrict our attention to one-dimensional MHD planar flow, which implies that
we are following the shock evolution along the jet axis. 
The fluid is described in terms of its density $\rho$, 
velocity $u$, thermal pressure $p$ and (transverse) magnetic 
field $B_y$. These quantities obey the standard MHD equations 
for conservation of mass, momentum, magnetic field and 
total energy, in the presence of radiative cooling represented by
the energy loss term ${\cal L}(T, \mbox{\boldmath${X}$})$ (energy per 
unit volume per unit time), which depends on the temperature $T$ and 
on the ionization state of the plasma, described by the vector of 
ionization fractions {\boldmath${X}$}. A discussion on the cooling function
follows in the next subsection.

A nonuniform preshock density that decreases away from the star is a reasonable 
assumption when dealing with an expanding jet, as suggested by observations. 
We note that shocks propagating into a stratified medium
tend to increase their amplitude when they find a decreasing preshock density.
We consider the following form for the preshock density:
\begin{equation}
\rho_0(x)=\rho_0 \frac{x_0^p}{x_0^p + x^p} \,.
\end{equation}
where $x$ is the spatial coordinate along the jet axis,
 $x_0$ sets the initial steepness of the density
function  (this affects the shock evolution even at larger distances) and
$1\leq p \leq 2$.

As in Paper I, the form of the initial  perturbation has been taken in
such a way as to keep the hydrodynamic Riemann
invariant $J_-=$constant. In this way a single forward shock is formed.  
In this scheme, the density perturbation is:
\begin{equation}
\rho=\left[ \frac{\gamma-1}{2 \sqrt{K\gamma}} (u-U_0)+\rho_0^{\frac{\gamma-1}{2}} 
\right]^{\frac{2}{\gamma-1}}  \;,
\label{eq:riem4}
\end{equation}
where $K$ is a constant derived from the single shock assumption, $\gamma = 5/3$
(the gas is monoatomic; molecules form at lower temperatures, while we are 
interested in the post-shock regions of high radiative emission), 
$U_0$ is the mean flow velocity
and the initial velocity perturbation is:
\begin{displaymath}
u(x)=\left\{ \begin{array}{ll}
u_0 [-(x-x_{\rm 1})^2 + 2 \sigma (x-x_{\rm 1})] &  \textrm{if \ $2 \sigma + x_{\rm 1}> x>x_{\rm 1}$ } \\
0 & \quad \; \textrm{otherwise}
\end{array} \right.
\end{displaymath}
where $u_0$ is the perturbation amplitude, $x_{\rm 1}$ is the initial 
coordinate of the perturbation and $\sigma$ is its half-width. However, the
exact shape of the perturbation is not crucial for the model.
In our calculations we take $x_1=10^{14}$ cm and $\sigma=2 
\times 10^{13}$ cm.

In carrying out the calculations we have set $U_0(x)=0$, i.e. 
the reference frame is that of the mean flow 
(that is, the steady jet flow on which the perturbation is set).  
To transforme the results back to the laboratory frame we have set $U_0=100$ km s$^{-1}$,
which is the bulk velocity we assume for the HH 30 jet ($U_0$ must not be mistaken for $u_0$, 
the initial velocity perturbation amplitude).

The boundary conditions assume free outflow at $x=0$ and $x=L$.
The extent of the computational domain, $L$, has been chosen 
sufficiently large to follow the shock evolution for 
$t \sim 15$ yrs and to avoid spurious interactions with the boundaries. 
For this reason, we adopt  $L = 4.5 \times 10^{15}$ cm.

The assumption of a monoatomic gas is justified in the case of HH30 and many other
YSO jets -- however, there are cases (the so-called ''molecular jets``) when this
assumption is not valid anymore, and molecular cooling must be taken into account.

\subsection{Radiative cooling}

In the current work we use two different approaches for the computation of the 
radiative cooling loss term ${\cal L}(T, \mbox{\boldmath${X}$})$. In a first
simplified model (see Paper I), {\boldmath${X}$} consists of the ionization fraction 
of H only, whereas a second, more accurate treatment is summarized below.

In the detailed cooling treatment, described in Te\c{s}ileanu et al. (\cite{TA08}),
28 additional evolutionary equations are solved for  
the non-equilibrium ionization fractions {\boldmath${X}$} (of H, He, C, 
N, O, Ne and S). 
The loss term accounts for energy lost in lines as well as in the ionization
and recombination processes. Line emissions include contributions  
from transitions of the 29 ion species.
The metal abundances are the ones adopted by Bacciotti et al. (\cite{BE99}),
in particular $N/H=1.1 \times 10^{-4}$, $O/H=6\times 10^{-4}$, $S/H=1.6 \times 10^{-5}$
and the abundance of $C$ is 10\% of the solar one.

To carry out comparisons between the observed and computed line intensity
ratios, we have determined the populations of the atomic levels relevant for
[SII], [NII] and [OI] emission, solving the excitation - de-excitation equilibrium
equations for five energy levels,  according to \cite{OF06}.
The line emissions reported in Bacciotti et al. (\cite{BE99}) for the HH 30 jet 
are: [SII]]$\lambda 6716 + \lambda 6731$, [NII]$\lambda 6548 + \lambda 6583$ 
and [OI]$\lambda 6300$,
and the corresponding intensity ratios [OI]/[NII] and [SII]/[OI].

\section{The case of the HH 30 jet}

To illustrate the astrophysical application of the above 
methodology, we will apply it to the numerical simulation
of some of the physical properties of the HH 30 jet. 

We have integrated the magneto-fluid equations employing the AMR method 
described previously with eight levels of refinement, using the PLUTO code
(Mignone et al. \cite{MA07}). 
As in Paper I, in order to obtain values directly comparable with observations, 
we have space averaged all post-shock quantities at each evolutionary time point:
\begin{equation}
\langle Q(t) \rangle= \frac{\int Q(x,t) \epsilon \{[SII](x,t)\} \ dx}
{ \int \epsilon \{[SII](x,t)\} \ dx}
\end{equation}
where $Q(x,t)$ is a physical quantity such as electron density or ionization fraction
or temperature.
This procedure is applied because by processing the observational data, we only have 
access to the physical parameters of the line-emitting regions, so this weighting is
implicit. Differently from Paper I, where we averaged the
line intensity ratios using the total line emissivity of the corresponding emitters as 
weighting functions, in the present paper we find the unweighted average 
of the line emissivity and calculate the ratio of the average lines afterward. 
This procedure does not
lead to results that differ qualitatively from the previous ones, but appears closer
to the actual observation process.
To calculate the line emissivities, the atomic transition rates and 
collision strengths adopted are the ones described in Te\c{s}ileanu et al. 
(\cite{TA08}). The collision strengths have been interpolated using a 
Lagrangian scheme to account for their temperature dependence. 

The main free parameters of the model are: the
particle density (at $x=0$) $n_0$, $x_0$, the velocity perturbation amplitude $u_0$ and 
initial transverse magnetic field $B_0$. Taking advantage of the peculiar efficiency of
the AMR method, we have widely explored this parameter space to find good agreement with the
observed line ratios. We have set the (uniform)
preshock temperature $T_0=1,000$ K and the initial ionization fraction $f_{\rm i}
 = 0.1\%$, i.e. due to the metal contribution only.

Both radiative cooling treatments were employed in the simulations, searching for the
best agreement of the model results with the observations for each one. A discussion 
of the differences in the results follows.

\subsection{Simplified cooling}\label{ss:raym}

The simplicity of this cooling model allows for rapid simulations, and thus the 
very efficient exploration of the parameter space. 
The root mean square deviations of the simulated line ratios with respect to
the observational ones were computed (a few examples are shown in Table \ref{tab:conv})
in order to quantitatively estimate how close each simulation is to observations.
The simulations presented in the table have base densities of $2\times 10^4$cm$^{-3}$ 
(n2e4) or $5\times 10^4$cm$^{-3}$ (n5e4), transverse magnetic field 300$\mu$G (b300) 
or 500$\mu$G (b500), and amplitude of the perturbation in velocity 50 (v50), 55 (v50) 
and 70km\,s$^{-1}$ (v70).

\begin{table}[h]
  \begin{center}
  \begin{tabular}{c|c|c}
    {\bf Simulation} & [OI]/[NII] & [SII]/[OI] \\
    \hline \hline
    n5e4b500v70 & 0.440 & 0.305   \\    
    n2e4b500v55 & 0.161 & 0.103   \\
    n5e4b500v55 & 0.121 & 0.143   \\
    n2e4b500v50 & 0.068 & 0.110   \\
    n2e4b300v50 & 0.063 & 0.108   \\
    \hline
  \end{tabular}
  \end{center}
  \caption{Root mean square deviations of line ratios with simplified cooling.}
  \label{tab:conv}
\end{table}

\begin{figure}
\resizebox{\hsize}{!}{\includegraphics{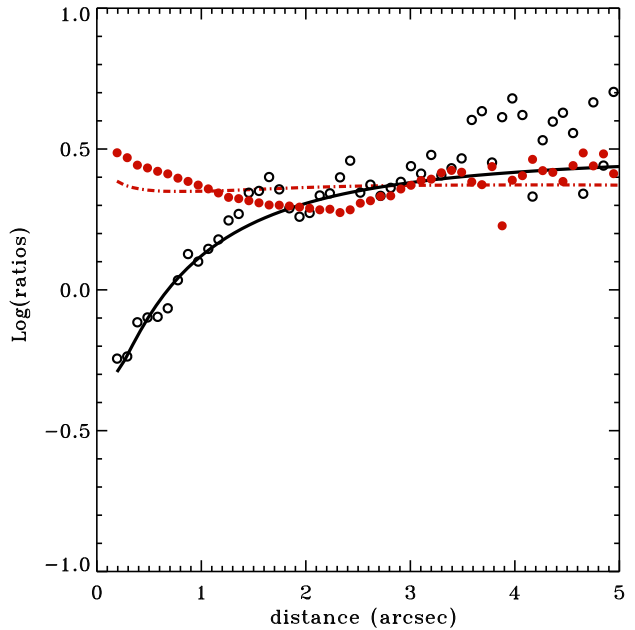}}
\caption{Intensity ratios of [SII]]$\lambda
6716 + \lambda 6731$, [NII]$\lambda 6548 + \lambda 6583$ and [OI]$\lambda 6300$ lines  
vs distance. Symbols represent the data from 
Bacciotti et al. (\cite{BE99}) and curves are the computed model curves 
(using the {\it simplified cooling} model):  [OI]/[NII] (bullets, 
dot-dashed line) and [SII]]/[OI] (circles, solid line). The model parameters are
$n_0=2 \times 10^4$ cm$^{-3}$, $B_0 =300 \mu$G, $x_0=0.^{\prime \prime}1$ and
$u_0=50$ km s$^{-1}$, with $U_0=100$ km s$^{-1}$ as the jet bulk velocity.
          }
\label{fig:rat}
\end{figure}

A combination of the parameters that yielded a reasonably good agreement with the observed
line intensity ratios, employing the simplified cooling model, was 
$n_0=2 \times 10^4$ cm$^{-3}$, $B_0 =300 \mu$G, $x_0=0.^{\prime \prime}1$ and
$u_0=50$ km s$^{-1}$. In Paper I for the DG Tau jet we set $p=2$, in the present
case a better agreement is obtained setting $p=1$, i.e. for a less steep decrease of the
preshock density with distance, i.e. a slower expansion rate for the jet as it propagates into
the ambient medium. This can be explained by the higher degree of collimation of the HH 30 jet. 
A comparison with Paper I shows also that $u_0$ is lower in this case 
with respect to the DG Tau parameters, i.e. $u_0=50$ km s$^{-1}$ instead of $u_0=70$ km s$^{-1}$; this
is reflected in the value of the ionization fraction, considerably lower for HH 30.
Moreover, we note
 a higher value of $B_0$ that must be assumed in the present case with respect to DG Tau.
The results are shown in Fig. \ref{fig:rat}, where we plot the 
observed line intensity ratios of [OI]/[NII] (bullets) and [SII]/[OI] (circles)
(as reported by Bacciotti et al. \cite{BE99})  and, superimposed, the corresponding 
model curves ($\langle$[OI]$\rangle$/$\langle$[NII]$\rangle$ dot-dashed line 
and $\langle$[SII]$\rangle$/$\langle$[OI]$\rangle$ solid line). 
Since the model proposed is extremely simple, we mainly aim, as
shown in Fig. \ref{fig:rat}, to obtain values of the intensity
ratios and trends with distance that are reasonably close to the ones observed,
and we cannot account for the knots of emission visible along the HH 30 jet, resulting in
ample spatial oscillations of the line intensity ratios, especially in the [OI]/[NII] ratio. 

Note also that a different choice of the metal abundances can 
shift the model lines of Fig. \ref{fig:rat} by some percent; e.g. \cite{Lo03} reports 
$N/H=0.802 \times 10^{-4}$, $O/H=5.81 \times 10^{-4}$ and $S/H=1.83 \times 10^{-5}$.

\begin{figure}
\resizebox{\hsize}{!}{\includegraphics{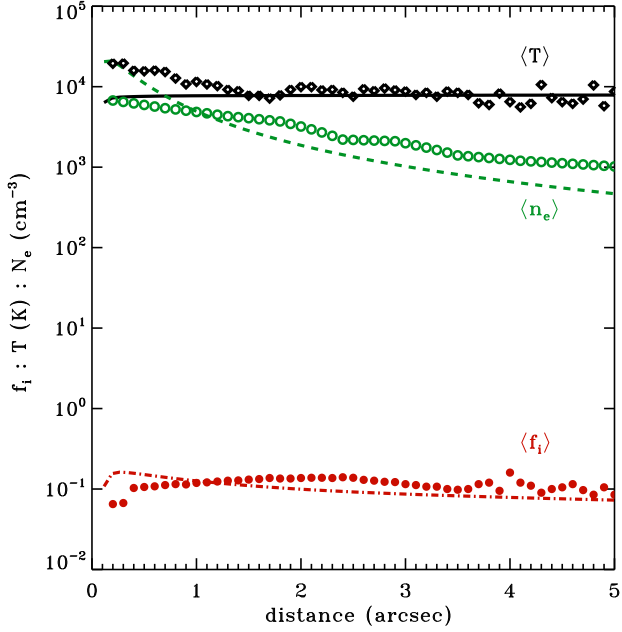}}
\caption{Average physical parameters of the jet derived from
observations following Bacciotti et al. (\cite{BE99}) (symbols) and from the model with {\it simplified cooling} (curves).
Temperature (diamonds, solid line), electron density (circles, dashed) and ionization
fraction (bullets, dot-dashed).
          }
\label{fig:par}
\end{figure}

In Fig. \ref{fig:par} we plot the physical parameters that are derived with a special 
diagnostic technique from observations
by Bacciotti et al. (\cite{BE99}) (symbols) compared to the outcome of the present model (lines).
We note that the electron density derived from the shock model
(dashed line) exceeds the one derived by Bacciotti et al. (\cite{BE99}) (circles) 
for distances below $\sim 1^{\prime \prime}$. As discussed in
Paper I, however, when the electron density is above the critical density, the ratio 
of observed [SII] lines, from which $n_e$ is derived, saturates, and
the diagnostic in this condition, as reported in the aforementioned paper,
 yields only lower limits to the electron density. The discrepancy in the
electron density affects the diagnostics of temperature (diamonds) 
 that differs from the model one (solid line).

\subsection{Detailed cooling}

The previous cooling model might be an oversimplified approximation in the case of 
moderately powerful shocks such as the ones we are dealing with here. This leads to
one of the motivations of this paper: a comparative study of the differences between and
advantages of each of the two approaches to radiative cooling treatment.

\begin{figure}
\resizebox{\hsize}{!}{\includegraphics{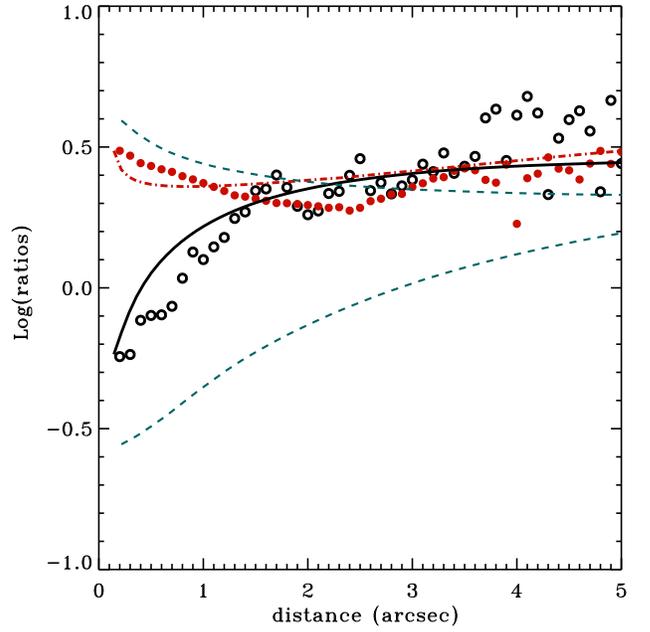}}
\caption{Intensity ratios of [SII]]$\lambda
6716 + \lambda 6731$, [NII]$\lambda 6548 + \lambda 6583$ and [OI]$\lambda 6300$ lines  
vs distance. Symbols represent the data from 
Bacciotti et al. (\cite{BE99}) and curves are the computed model curves (with {\it detailed cooling}
treatment):  [OI]/[NII] (bullets, 
dot-dashed line) and [SII]]/[OI] (circles, solid line). The model parameters are
$n_0=5 \times 10^4$ cm$^{-3}$, $B_0 =500 \mu$G, $x_0=0.^{\prime \prime}1$ and
$u_0=55$ km s$^{-1}$, with $U_0=95$ km s$^{-1}$ as the jet bulk velocity.
The dashed, lighter lines are the results obtained with the simplified cooling
model for the same parameter set.
          }
\label{fig:rat_m}
\end{figure}

\begin{figure}
\resizebox{\hsize}{!}{\includegraphics{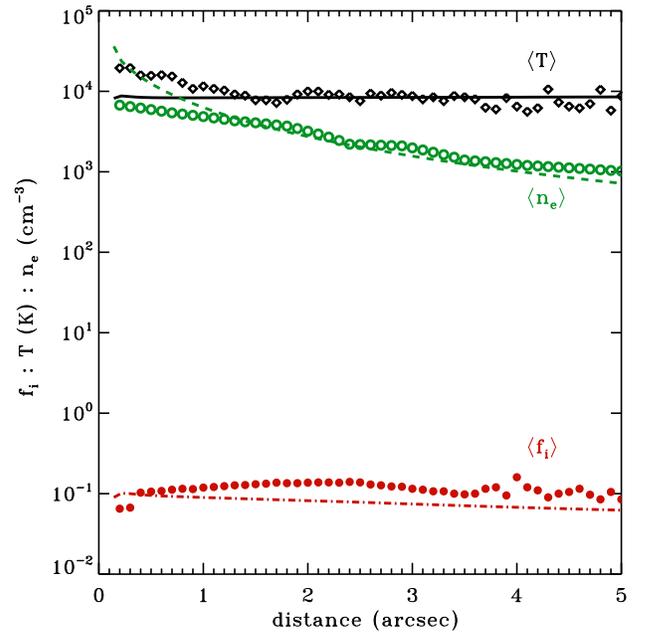}}
\caption{Average physical parameters of the jet derived from
observations following Bacciotti et al. (\cite{BE99}) (symbols) and from the model 
with {\it detailed cooling} (curves). Temperature (diamonds, solid line), electron 
density (circles, dashed) and total ionization fraction (bullets, dot-dashed).
          }
\label{fig:par_m}
\end{figure}

Employing the detailed cooling model, we have explored the parameter space for the 
same setup as in the previous case. 
The same method of estimating the root mean square deviation with respect to observations
was employed. The best agreement with the observed
line intensity ratios was obtained for the following parameters: 
$n_0=5 \times 10^4$ cm$^{-3}$, $B_0 =500 \mu$G, $x_0=0.^{\prime \prime}1$,
$u_0=55$ km s$^{-1}$ and $p=1$. The resulting line ratios are plotted in 
Fig. \ref{fig:rat_m} and the evolution of the physical parameters in
Fig. \ref{fig:par_m}. In this scheme, the full ionization state of the plasma is 
computed at each step, allowing for non-equilibrium states. This of course also has 
a serious impact on the simulation speed, because of the supplementary 
equations in the system.


The spatial profile of the physical parameters is similar to the one obtained 
in the best agreement case employing the simplified cooling. The same considerations
of the underestimation of the electron density close to the source apply.

By applying the set of simulation parameters resulting from the simulations with
detailed cooling to the setup employing simplified cooling, important differences 
appear. Due to the higher overall density of the plasma and the much simpler chemistry
in the simplified cooling case, the electron density remains at values above the 
ones derived from observations by a factor of 3 along the evolution of the shock.
This is also reflected in the line ratios that depart from the observed ones by a factor
of 2 for the $[OI]/[NII]$ and 20\% for the $[SII]/[OI]$ ratios, triggered by the
higher maximum temperatures attained.

\subsection{On the importance of numerical resolution}

It is important to understand the resolution needed for these simulations, 
before moving to 2D configurations (Te\c{s}ileanu et al. \cite{TA09}). 
A rough estimation of the resolution needed in order to resolve the post-shock zone 
can be done as follows: for a maximum peak temperature of about $10^5$K and densities of 
$\lesssim 10^5$cm$^{-3}$, the internal energy density of the gas is of the order of 
$10^{-6}$\,erg\,cm$^{-3}$.
According to previous computations (see Te\c{s}ileanu et al. \cite{TA08}), the maximum 
energy loss rate in radiative cooling processes in these conditions is $\sim 10^{-12}$\,erg\,cm$^{-3}$s$^{-1}$.
Allowing for a maximum relative change in internal energy of 1\% in a timestep, the size of
the timestep is restricted to about $10^4$s, which combined with the jet velocity of a 
few hundreds km\,s$^{-1}$ leads to a space scale of $\sim 10^{11}$cm.
Even higher resolutions, of a few times $10^{10}$cm, are needed to accurately resolve the 
ionization/recombination processes (with timescales of $\lesssim 10^3$s). For the setups used 
in this work, this spatial resolution is attained with 8 levels of refinement.

We have studied the effect of increasing resolution on the line ratios and found it to be 
important up to a saturation value above which the line ratios do not change significantly.
For this, we have computed the root mean square deviations between the simulated values
of the line ratios with an increasing number of refinement levels. The results are presented in 
Table \ref{tab:res}, where the first column shows the number of refinement levels in the simulations
between which the RMS deviations were computed. Each refinement level doubles the 
resolution. Three configurations are presented: one with base density N=$2\times 10^4$cm$^{-3}$, 
transverse magnetic field 300$\mu$G, perturbation amplitude in velocity 50km\,s$^{-1}$; the second 
one has N=$5\times 10^4$cm$^{-3}$, B=100$\mu$G, velocity perturbation 40km\,s$^{-1}$; the third one 
has N=$5\times 10^4$cm$^{-3}$, B=500$\mu$G, velocity perturbation 55km\,s$^{-1}$. The convergence 
of the results can be inferred from these data.

\begin{table}[h]
  \begin{center}
  \begin{tabular}{c|c|c|c|c|c|c}
    {\bf Sim.} & \multicolumn{2}{|c|}{n2e4b300v50} &  \multicolumn{2}{|c}{n5e4b100v40} & 
    									 \multicolumn{2}{|c}{n5e4b500v55} \\
    \hline
        ref.levs.     &  R1 & R2 &  R1 & R2 &  R1 & R2 \\
    \hline \hline
    base -- 2 & 0.1378 & 0.0319 & 0.1240 & 0.0477 & 0.1150 & 0.0477 \\
    2 -- 4    & 0.0512 & 0.0171 & 0.0883 & 0.0534 & 0.0884 & 0.0499 \\
    4 -- 6    & 0.0141 & 0.0816 & 0.0425 & 0.0418 & 0.0589 & 0.0509 \\
    6 -- 8    & 0.0110 & 0.0141 & 0.0088 & 0.0145 & 0.0256 & 0.0359 \\
    8 -- 10   & 0.0004 & 0.0008 & 0.0017 & 0.0023 & 0.0033 & 0.0084 \\
    \hline
  \end{tabular}
  \end{center}
  \caption{ Root mean square deviations of line ratios (R1 is [OI]/[NII], R2 is [SII]/[OI])
     obtained with simulations of increasing resolution.}
  \label{tab:res}
\end{table}

\begin{figure}
\resizebox{\hsize}{!}{\includegraphics{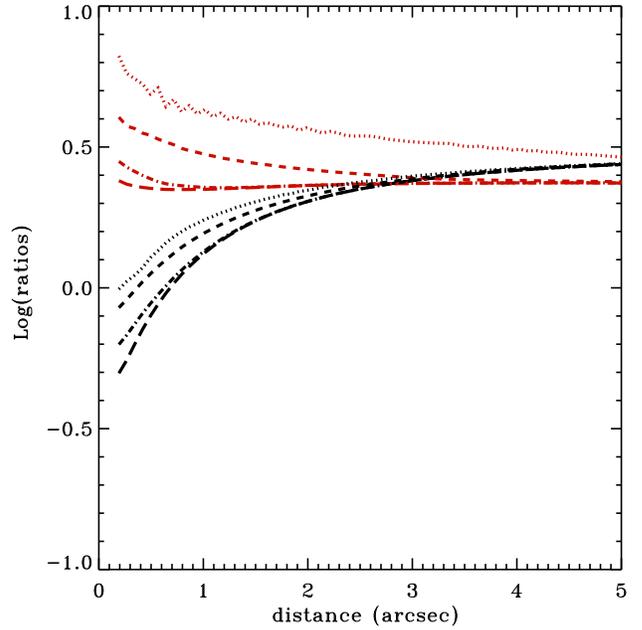}}
\caption{Effect of the simulation resolution on the line ratios of [SII]]$\lambda
6716 + \lambda 6731$, [NII]$\lambda 6548 + \lambda 6583$ and [OI]$\lambda 6300$ lines,
in the case of {\it simplified cooling}. The setup is identical to the one in Fig. \ref{fig:rat},
with the four line sets representing results from simulations with 10 levels of refinement
(long-dashed lines), 6 levels of refinement (dash-dotted lines), 3 levels of refinement 
(dashed lines) and without refinement at all (dotted lines), respectively.
  }
\label{fig:resol}
\end{figure}

Presented in Fig. \ref{fig:resol} are four cases of increasing resolution for the
first configuration in Table \ref{tab:res} (n2e4b300v50): one employing
only the base grid (1\,536 zones), a second one with three levels of refinement 
(equivalent resolution 12\,288 cells), one using six levels
of refinement (equivalent resolution 98\,304 cells), and one with 10 levels of 
refinement (equivalent resolution 1\,572\,864 cells). It can be seen that 
the line ratios converge to the results of Fig. \ref{fig:rat}, obtained in a simulation
with 8 levels of refinement. The simulations presented in Fig. \ref{fig:resol} use
the simplified cooling model, but similar effects were obtained when using the detailed
cooling model. In Fig. \ref{fig:resol}, the line ratios obtained employing 8 levels of
refinement almost perfectly overlapped the ones obtained with 10 levels of refinement,
so they were not shown.

The physical spatial resolution in a simulation with 8 levels of refinement in the present 
setup is about $1.2\times 10^{10}$cm, and should be enough to adequately resolve the post-shock zone, 
with its rapid variation of physical parameters.  The resolution for 6 levels of 
refinement is $\sim$5$\times 10^{10}$cm, it drops to $4\times 10^{11}$cm for three levels
of refinement and to $3.25\times 10^{12}$cm at the base grid resolution.

The line ratios are very sensitive to resolution
variations, making it important to reach an adequate (high) resolution for reliable results.
This is a critical aspect for future 2D simulations, that will have to employ AMR to
reach these extremely high resolutions with the available computational power.

\begin{figure}
\begin{center}
\resizebox{6.0cm}{!}{\includegraphics{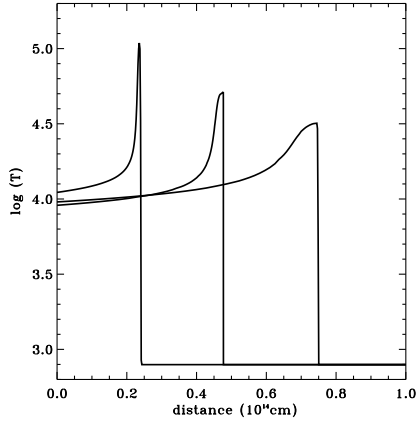}}
\caption{Three simulations of shockwaves with density N=$2\times 10^4$cm$^{-3}$, transverse magnetic fields
     300$\mu$G, and perturbation amplitudes in velocity of (from left to right) 70, 50 and 40\,km\,s$^{-1}$, 
     respectively, at the same evolutionary point. Logarithmic plots of temperature.
  }
\label{fig:resol2}
\end{center}
\end{figure}

\begin{figure}
\begin{center}
\resizebox{6.0cm}{!}{\includegraphics{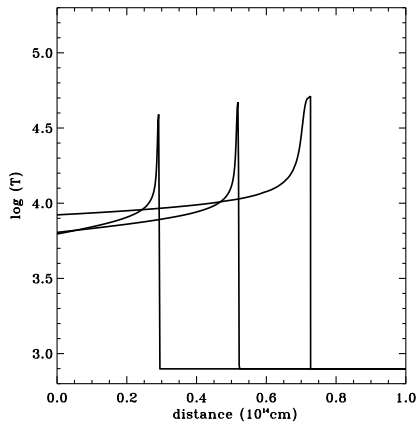}}
\caption{Three simulations of shockwaves with transverse magnetic fields 300$\mu$G, perturbation 
     amplitude in velocity of 50\,km\,s$^{-1}$, and base densities of (from left to right) 
     $8$,  $5$, and $2\times 10^4$cm$^{-3}$, respectively, at the same evolutionary point. 
     Logarithmic plots of temperature.
}
\label{fig:resol3}
\end{center}
\end{figure}

In order to see how the resolution requirements depend on the parameters of the shock, 
in Fig. \ref{fig:resol2} we present the evolution of the temperature at the same age
for three shockwaves with different amplitudes of the initial perturbation (decreasing
from left to right). 
The $x$ axis does not represent the actual position of the shocks but only a distance
scale, the front of the shockwaves being plotted close to each other for convenience.
We see that with the increase of the perturbation amplitude (and thus the shock strength),
the evolution after the shock steepens. In the same way, in Fig. \ref{fig:resol3} three 
shockwaves with different initial densities are plotted, with decreasing density from 
left to right. The evolution of the temperature after the shock steepens when 
increasing the density. Typically, as the dependence steepens, the resolution needed
increases. All the cited plots were made for the
same average velocity of the underlying jets. When decreasing this velocity, the spatial
resolution needed for an accurate analysis of the post-shock zone increases.

\subsection{Comparison and discussion}

The parameter sets for the best agreement obtained 
in the previous paragraphs were not the only ones that well interpret the observations, 
but other possible ``successful" combinations do not depart from these values 
of the parameters. As an example, setting $U_0=95$ km s$^{-1}$ for 
the bulk jet velocity, we obtained model curves very similar to those of 
Figs. \ref{fig:rat_m} and \ref{fig:par_m} with the same $p=1$, $x_0=0.^{\prime \prime}1$ 
but with $n_0=4 \times 10^4$ cm$^{-3}$, $B_0 =300 \mu$G, and $u_0=58$ km s$^{-1}$. The
same considerations apply for the simplified cooling simulations.

The jet bulk speeds $U_0$ were chosen in such a way that the total shock speed in
the rest reference frame would be approximately equal to 150km\,s$^{-1}$ in all cases --
this being an approximate average knot speed derived from observations of Doppler shifts
and proper motions (Hartigan \& Morse \cite{HM07}).

Comparing the results obtained with the two approaches to radiative cooling, 
identical or similar values of the model parameters can be inferred (Table \ref{tab:comp}).
Parameters such as velocity amplitude of the perturbation (determining the
shock strength) and transverse magnetic field (resulting in shock compression)
are similar in the two cases. The total density of the jet is however one of 
the most difficult parameters to estimate from observations, and the different
results in our two cases come from the different evolution of the total 
radiative losses and ionization in the post-shock zone.

\begin{table}[h]
  \begin{center}
  \begin{tabular}{c|c|c}
    Parameter  & Simplified & Detailed  \\
    \hline \hline
    $n_0$(cm$^{-3}$)      & 2$\times 10^4$ & 5$\times 10^4$    \\
    $B_0$($\mu$G)         &  300  &  500   \\
    $u_0$(km\,s$^{-1}$)   &   50  &  55    \\
    $U_0$(km\,s$^{-1}$)   &  100  &  100   \\
    $f_{\rm i}$(\%)       &  0.1  &  0.1   \\
    $T_0$(K)              &  1000 &  1000  \\
    \hline
  \end{tabular}
  \end{center}
  \caption{Comparison of the model parameters that gave the best agreement 
     with the observations of HH30. }
  \label{tab:comp}
\end{table}

The simplified radiative cooling treatment appear to be suited for a rapid exploration
of the parameter space, giving results (parameters of the model) close to the ones
obtained with more sophisticated cooling.

\section{Conclusions and summary}

We have demonstrated in previous works that numerical integration of 
time-dependent radiating (shocked) flows can substantially benefit from the 
employment of adaptive mesh refinement methods, and the present work 
supports this conclusion.
By computing the cooling time scale according to the algorithm described in
Mignone et al. (\cite{MA09}), one could take full advantage of the time 
adaptation process.
The use of this method allows us to tackle simulations of radiating shocks 
in 2D as they propagate along a cylindrical supersonic jet. 

Our results applied to the DG Tau jet (Paper I) as well as the HH 30 jet (present paper) 
revealed a good agreement between the observed and calculated line intensity 
ratios of different transitions and for different elements (Figs. \ref{fig:rat} and \ref{fig:rat_m}), 
and also of derived physical parameters, including the ionization fraction 
(Figs. \ref{fig:par} and \ref{fig:par_m}). 

An accurate treatment of the radiative losses allows us to more realistically compute 
the full ionization state of the plasma, and thus accurate line emission for
each ion, in the approximation of a five-level atom. This, however, requires significant
additional computing power that is not always available. An alternative is the use of
the simplified cooling model, following only the ionization state of hydrogen, that 
proved to give good indications on the perturbation amplitude and transverse magnetic 
field of our model. This simpler but much faster strategy may be used for a preliminary 
exploration of the parameter space, being followed by simulations with detailed cooling 
only in the areas of interest. 
Both cooling models were discussed and compared in Te\c{s}ileanu et al. (\cite{TA08}),
and in the present work we extended the comparison to line ratio diagnostics and 
application to a real HH object.

Grid resolution proved to be a critical parameter for the line ratios estimations 
from MHD simulations. The physical dimensions of the computational zones have to not exceed 
about $10^{11}$ cm in order to accurately capture the evolution of the
physical parameters and emission properties in the post-shock zones. This is a
crucial aspect for future 2D simulations that will have to fulfill these 
requirements. This result can be generalized for the typical conditions encountered 
in shocks traveling through protostellar jets, as the variability in these conditions
could not change the numbers by an order of magnitude. Weaker shocks (shocks evolving from
smaller initial perturbations) or lighter shocks (lower initial density) will be less stringent 
on the resolution requirements, while stronger and/or denser shocks will add a factor to
the present results.

We can summarize the general picture as follows: i) when one is interested in the 
distribution in space of the jet physical parameters such as density, temperature, etc., 
the details of the assumed cooling function matter very little (Te\c{s}ileanu et al. \cite{TA08}) provided
the numerical resolution suffices to minimize numerical dissipation effects; ii) when one
considers the detailed shock structure, numerical resolution is more important, but
less so for the behavior of the integrated emission line luminosities (\cite{Raga07});
iii) when one discusses the line intensity ratios, a finer resolution must be achieved,
while the details of the cooling function have effects that differ according to the
line considered, in the present case from about 10\% up to a factor of two.

\begin{acknowledgements} 
We acknowledge the Italian MIUR for financial support, grants No. 2002.028843 and No. 2004.025227.
The present work was supported in part by the European Community Marie Curie
Actions - Human resource and mobility within the JETSET network under
contract MRTN-CT-2004 005592. OT acknowledges the support from the Romanian 
Research Agency CNCSIS, contract number RP-4/2009. We thank the anonymous referee for valuable
observations and suggestions.

\end{acknowledgements} 

{}

\end{document}